\PassOptionsToPackage{unicode}{hyperref}
\PassOptionsToPackage{hyphens}{url}
\PassOptionsToPackage{dvipsnames,svgnames,x11names}{xcolor}
\documentclass[
]{article}
\usepackage{amsmath,amssymb}
\usepackage{lmodern}
\usepackage{iftex}
\ifPDFTeX
  \usepackage[T1]{fontenc}
  \usepackage[utf8]{inputenc}
  \usepackage{textcomp} % provide euro and other symbols
\else % if luatex or xetex
  \usepackage{unicode-math}
  \defaultfontfeatures{Scale=MatchLowercase}
  \defaultfontfeatures[\rmfamily]{Ligatures=TeX,Scale=1}
\fi
\IfFileExists{upquote.sty}{\usepackage{upquote}}{}
\IfFileExists{microtype.sty}{% use microtype if available
  \usepackage[]{microtype}
  \UseMicrotypeSet[protrusion]{basicmath} % disable protrusion for tt fonts
}{}
\makeatletter
\@ifundefined{KOMAClassName}{% if non-KOMA class
  \IfFileExists{parskip.sty}{%
    \usepackage{parskip}
  }{% else
    \setlength{\parindent}{0pt}
    \setlength{\parskip}{6pt plus 2pt minus 1pt}}
}{% if KOMA class
  \KOMAoptions{parskip=half}}
\makeatother
\usepackage{xcolor}
\newlength{\cslhangindent}
\newlength{\csllabelwidth}
\newlength{\cslentryspacingunit} % times entry-spacing
\newenvironment{CSLReferences}[2] % #1 hanging-ident, #2 entry spacing
 {% don't indent paragraphs
  \setlength{\parindent}{0pt}
  \ifodd #1
  \let\oldpar\par
  \def\par{\hangindent=\cslhangindent\oldpar}
  \fi
  \setlength{\parskip}{#2\cslentryspacingunit}
 }%
 {}
\usepackage{calc}

\ifLuaTeX
\usepackage[bidi=basic]{babel}
\else
\usepackage[bidi=default]{babel}
\fi
\babelprovide[main,import]{american}
\def\languageshorthands#1{}
\ifLuaTeX
  \usepackage{selnolig}  % disable illegal ligatures
\fi
\IfFileExists{bookmark.sty}{\usepackage{bookmark}}{\usepackage{hyperref}}
\IfFileExists{xurl.sty}{\usepackage{xurl}}{} % add URL line breaks if available
\hypersetup{
  pdftitle={SLEPLET: Slepian Scale-Discretised Wavelets in Python},
  pdfauthor={Patrick J. Roddy},
  pdflang={en-US},
  colorlinks=true,
  linkcolor={Maroon},
  filecolor={Maroon},
  citecolor={Blue},
  urlcolor={Blue},
  pdfcreator={LaTeX via pandoc}}

\title{SLEPLET: Slepian Scale-Discretised Wavelets in Python}

\usepackage[affil-it]{authblk}
\usepackage{orcidlink}
\author[1%
  ]{Patrick J. Roddy%
    \,\orcidlink{0000-0002-6271-1700}\,%
    }

\affil[1]{Advanced Research Computing, University College London, UK}
\date{20\textsuperscript{th} April 2023}

\begin{document}
\maketitle

\hypertarget{summary}{%
\section{Summary}\label{summary}}

Wavelets are widely used in various disciplines to analyse signals both
in space and scale. Whilst many fields measure data on manifolds (i.e.,
the sphere), often data are only observed on a partial region of the
manifold. Wavelets are a typical approach to data of this form, but the
wavelet coefficients that overlap with the boundary become contaminated
and must be removed for accurate analysis. Another approach is to
estimate the region of missing data and to use existing whole-manifold
methods for analysis. However, both approaches introduce uncertainty
into any analysis. Slepian wavelets enable one to work directly with
only the data present, thus avoiding the problems discussed above.
Applications of Slepian wavelets to areas of research measuring data on
the partial sphere include gravitational/magnetic fields in geodesy,
ground-based measurements in astronomy, measurements of whole-planet
properties in planetary science, geomagnetism of the Earth, and cosmic
microwave background analyses.

\hypertarget{statement-of-need}{%
\section{Statement of Need}\label{statement-of-need}}

Many fields in science and engineering measure data that inherently live
on non-Euclidean geometries, such as the sphere. Techniques developed in
the Euclidean setting must be extended to other geometries. Due to
recent interest in geometric deep learning, analogues of Euclidean
techniques must also handle general manifolds or graphs. Often, data are
only observed over partial regions of manifolds, and thus standard
whole-manifold techniques may not yield accurate predictions. Slepian
wavelets are designed for datasets like these. Slepian wavelets are
built upon the eigenfunctions of the Slepian concentration problem of
the manifold (\protect\hyperlink{ref-Landau1961}{Landau \& Pollak,
1961}, \protect\hyperlink{ref-Landau1962}{1962};
\protect\hyperlink{ref-Slepian1961}{Slepian \& Pollak, 1961}): a set of
bandlimited functions that are maximally concentrated within a given
region. Wavelets are constructed through a tiling of the Slepian
harmonic line by leveraging the existing scale-discretised framework
(\protect\hyperlink{ref-Leistedt2013}{Leistedt, B. et al., 2013};
\protect\hyperlink{ref-Wiaux2008}{Wiaux et al., 2008}). Whilst these
wavelets were inspired by spherical datasets, like in cosmology, the
wavelet construction may be utilised for manifold or graph data.

To the author's knowledge, there is no public software that allows one
to compute Slepian wavelets (or a similar approach) on the sphere or
general manifolds/meshes. \texttt{SHTools}
(\protect\hyperlink{ref-Wieczorek2018}{Wieczorek \& Meschede, 2018}) is
a \texttt{Python} code used for spherical harmonic transforms, which
allows one to compute the Slepian functions of the spherical polar cap
(\protect\hyperlink{ref-Simons2006}{Frederik J. Simons et al., 2006}). A
series of \texttt{MATLAB} scripts exist in \texttt{slepian\_alpha}
(\protect\hyperlink{ref-Simons2020}{Frederik J. Simons et al., 2020}),
which permits the calculation of the Slepian functions on the sphere.
However, these scripts are very specialised and hard to generalise.

\texttt{SLEPLET} (\protect\hyperlink{ref-Roddy2023a}{Roddy, 2023}) is a
Python package for the construction of Slepian wavelets in the spherical
and manifold (via meshes) settings. In contrast to the aforementioned
codes, \texttt{SLEPLET} handles any spherical region as well as the
general manifold setting. The API is documented and easily extendible,
designed in an object-orientated manner. Upon installation,
\texttt{SLEPLET} comes with two command line interfaces -
\texttt{sphere} and \texttt{mesh} - that allow one to easily generate
plots on the sphere and a set of meshes using \texttt{plotly}. Whilst
these scripts are the primary intended use, \texttt{SLEPLET} may be used
directly to generate the Slepian coefficients in the spherical/manifold
setting and use methods to convert these into real space for
visualisation or other intended purposes. The construction of the
sifting convolution (\protect\hyperlink{ref-Roddy2021}{Roddy \& McEwen,
2021}) was required to create Slepian wavelets. As a result, there are
also many examples of functions on the sphere in harmonic space (rather
than Slepian) that were used to demonstrate its effectiveness.
\texttt{SLEPLET} has been used in the development of
(\protect\hyperlink{ref-Roddy2022a}{Roddy, 2022};
\protect\hyperlink{ref-Roddy2021}{Roddy \& McEwen, 2021},
\protect\hyperlink{ref-Roddy2022}{2022},
\protect\hyperlink{ref-Roddy2023}{2023}).

Whilst Slepian wavelets may be trivially computed from a set of Slepian
functions, the computation of the spherical Slepian functions themselves
are computationally complex, where the matrix scales as
\(\mathcal{O}(L^{4})\). Although symmetries of this matrix and the
spherical harmonics have been exploited, filling in this matrix is
inherently slow due to the many integrals performed. The matrix is
filled in parallel in \texttt{Python} using \texttt{concurrent.futures},
and the spherical harmonic transforms are computed in \texttt{C} using
\texttt{SSHT}. This may be sped up further by utilising the new
\texttt{ducc0} backend for \texttt{SSHT}, which may allow for a
multithreaded solution. Ultimately, the eigenproblem must be solved to
compute the Slepian functions, requiring sophisticated algorithms to
balance speed and accuracy. Therefore, to work with high-resolution data
such as these, one requires high-performance computing methods on
supercomputers with massive memory and storage. To this end, Slepian
wavelets may be exploited at present at low resolutions, but further
work is required for them to be fully scalable.

\hypertarget{acknowledgements}{%
\section{Acknowledgements}\label{acknowledgements}}

The author would like to thank Jason D. McEwen for his advice and
guidance on the mathematics behind \texttt{SLEPLET}. Further, the author
would like to thank Zubair Khalid for providing his \texttt{MATLAB}
implementation to compute the Slepian functions of a polar cap region,
as well as the formulation for a limited colatitude-longitude region
(\protect\hyperlink{ref-Bates2017}{Bates et al., 2017}).
\texttt{SLEPLET} makes use of several libraries the author would like to
acknowledge, in particular, \texttt{libigl}
(\protect\hyperlink{ref-Libigl2017}{Jacobson \& Panozzo, 2017}),
\texttt{NumPy} (\protect\hyperlink{ref-Harris2020}{Harris et al.,
2020}), \texttt{plotly} (\protect\hyperlink{ref-Plotly2015}{Inc.,
2015}), \texttt{SSHT} (\protect\hyperlink{ref-McEwen2011}{McEwen \&
Wiaux, 2011}), \texttt{S2LET}
(\protect\hyperlink{ref-Leistedt2013}{Leistedt, B. et al., 2013}).

\hypertarget{references}{%
\section*{References}\label{references}}
\addcontentsline{toc}{section}{References}

\hypertarget{refs}{}
\begin{CSLReferences}{1}{0}
\leavevmode\vadjust pre{\hypertarget{ref-Bates2017}{}}%
Bates, A. P., Khalid, Z., \& Kennedy, R. A. (2017). Slepian
spatial-spectral concentration problem on the sphere: Analytical
formulation for limited colatitude-longitude spatial region. \emph{IEEE
Transactions on Signal Processing}, \emph{65}(6), 1527--1537.
\url{https://doi.org/10.1109/TSP.2016.2646668}

\leavevmode\vadjust pre{\hypertarget{ref-Harris2020}{}}%
Harris, C. R., Millman, K. J., Walt, S. J. van der, Gommers, R.,
Virtanen, P., Cournapeau, D., Wieser, E., Taylor, J., Berg, S., Smith,
N. J., Kern, R., Picus, M., Hoyer, S., Kerkwijk, M. H. van, Brett, M.,
Haldane, A., Río, J. F. del, Wiebe, M., Peterson, P., \ldots{} Oliphant,
T. E. (2020). Array programming with {NumPy}. \emph{Nature},
\emph{585}(7825), 357--362.
\url{https://doi.org/10.1038/s41586-020-2649-2}

\leavevmode\vadjust pre{\hypertarget{ref-Plotly2015}{}}%
Inc., P. T. (2015). \emph{Collaborative data science}.
\url{https://plot.ly}

\leavevmode\vadjust pre{\hypertarget{ref-Libigl2017}{}}%
Jacobson, A., \& Panozzo, D. (2017). Libigl: Prototyping geometry
processing research in {C}++. \emph{SIGGRAPH Asia 2017 Courses}.
\url{https://doi.org/10.1145/3134472.3134497}

\leavevmode\vadjust pre{\hypertarget{ref-Landau1961}{}}%
Landau, H. J., \& Pollak, H. O. (1961). Prolate spheroidal wave
functions, {F}ourier analysis and uncertainty -- {II}. \emph{The Bell
System Technical Journal}, \emph{40}(1), 65--84.
\url{https://doi.org/10.1002/j.1538-7305.1961.tb03977.x}

\leavevmode\vadjust pre{\hypertarget{ref-Landau1962}{}}%
Landau, H. J., \& Pollak, H. O. (1962). Prolate spheroidal wave
functions, {F}ourier analysis and uncertainty -- {III}: The dimension of
the space of essentially time- and band-limited signals. \emph{The Bell
System Technical Journal}, \emph{41}(4), 1295--1336.
\url{https://doi.org/10.1002/j.1538-7305.1962.tb03279.x}

\leavevmode\vadjust pre{\hypertarget{ref-Leistedt2013}{}}%
Leistedt, B., McEwen, J. D., Vandergheynst, P., \& Wiaux, Y. (2013).
S2LET: A code to perform fast wavelet analysis on the sphere.
\emph{Astronomy \& Astrophysics}, \emph{558}, A128.
\url{https://doi.org/10.1051/0004-6361/201220729}

\leavevmode\vadjust pre{\hypertarget{ref-McEwen2011}{}}%
McEwen, J. D., \& Wiaux, Y. (2011). A novel sampling theorem on the
sphere. \emph{IEEE Transactions on Signal Processing}, \emph{59}(12),
5876--5887. \url{https://doi.org/10.1109/TSP.2011.2166394}

\leavevmode\vadjust pre{\hypertarget{ref-Roddy2022a}{}}%
Roddy, P. J. (2022). \emph{Slepian wavelets for the analysis of
incomplete data on manifolds} {[}PhD thesis, UCL (University College
London){]}. \url{https://paddyroddy.github.io/thesis}

\leavevmode\vadjust pre{\hypertarget{ref-Roddy2023a}{}}%
Roddy, P. J. (2023). \emph{SLEPLET: {S}lepian scale-discretised wavelets
in {P}ython}. \url{https://doi.org/10.5281/zenodo.7268074}

\leavevmode\vadjust pre{\hypertarget{ref-Roddy2021}{}}%
Roddy, P. J., \& McEwen, J. D. (2021). Sifting convolution on the
sphere. \emph{IEEE Signal Processing Letters}, \emph{28}, 304--308.
\url{https://doi.org/10.1109/LSP.2021.3050961}

\leavevmode\vadjust pre{\hypertarget{ref-Roddy2022}{}}%
Roddy, P. J., \& McEwen, J. D. (2022). Slepian scale-discretised
wavelets on the sphere. \emph{IEEE Transactions on Signal Processing},
\emph{70}, 6142--6153. \url{https://doi.org/10.1109/TSP.2022.3233309}

\leavevmode\vadjust pre{\hypertarget{ref-Roddy2023}{}}%
Roddy, P. J., \& McEwen, J. D. (2023). \emph{Slepian scale-discretised
wavelets on manifolds}. arXiv. \url{https://arxiv.org/abs/2302.06006}

\leavevmode\vadjust pre{\hypertarget{ref-Simons2006}{}}%
Simons, Frederik J., Dahlen, F. A., \& Wieczorek, M. A. (2006).
Spatiospectral concentration on a sphere. \emph{SIAM Review},
\emph{48}(3), 504--536. \url{https://doi.org/10.1137/S0036144504445765}

\leavevmode\vadjust pre{\hypertarget{ref-Simons2020}{}}%
Simons, Frederik J., Harig, C., Plattner, A., Hippel, M. von, \& Albert.
(2020). \emph{{slepian\_alpha}}.
\url{https://doi.org/10.5281/zenodo.4085210}

\leavevmode\vadjust pre{\hypertarget{ref-Slepian1961}{}}%
Slepian, D., \& Pollak, H. O. (1961). Prolate spheroidal wave functions,
{F}ourier analysis and uncertainty -- {I}. \emph{The Bell System
Technical Journal}, \emph{40}(1), 43--63.
\url{https://doi.org/10.1002/j.1538-7305.1961.tb03976.x}

\leavevmode\vadjust pre{\hypertarget{ref-Wiaux2008}{}}%
Wiaux, Y., McEwen, J. D., Vandergheynst, P., \& Blanc, O. (2008). Exact
reconstruction with directional wavelets on the sphere. \emph{Monthly
Notices of the Royal Astronomical Society}, \emph{388}(2), 770--788.
\url{https://doi.org/10.1111/j.1365-2966.2008.13448.x}

\leavevmode\vadjust pre{\hypertarget{ref-Wieczorek2018}{}}%
Wieczorek, M. A., \& Meschede, M. (2018). {SHTools}: Tools for working
with spherical harmonics. \emph{Geochemistry, Geophysics, Geosystems},
\emph{19}(8), 2574--2592. \url{https://doi.org/10.1029/2018GC007529}

\end{CSLReferences}

\end{document}